\documentclass[aps,floats,twocolumn,showpacs]{revtex4}
\usepackage[cp1251]{inputenc}
\usepackage[english]{babel}
\usepackage{amssymb}
\usepackage{amsmath}
\usepackage{graphicx}
 \def\ds{\displaystyle}
 \def\bc{\begin{center}}          \def\ec{\end{center}}
 \allowdisplaybreaks

\begin{document}
 \title{Two-dimensional simulations of nonlinear beam-plasma interaction in isotropic and magnetized plasmas}
 \author{I.V.{\,}Timofeev}
 \affiliation{Budker Institute of Nuclear Physics SB RAS, 630090, Novosibirsk, Russia \\
 Novosibirsk State University, 630090, Novosibirsk, Russia}
 \begin{abstract}

Nonlinear interaction of a low density electron beam with a uniform
plasma is studied using two-dimensional particle-in-cell (PIC)
simulations. We focus on  formation of coherent phase space
structures in the case, when a wide two-dimensional wave spectrum is
driven unstable, and we also study how nonlinear evolution of these
structures is affected by the external magnetic field. In the case
of isotropic plasma, nonlinear buildup of filamentation modes due to
the combined effects of two-stream and oblique instabilities is
found to exist and growth mechanisms of secondary instabilities
destroying the BGK--type nonlinear wave are identified. In the weak
magnetic field, the energy of beam-excited plasma waves at the
nonlinear stage of beam-plasma interaction goes predominantly to the
short-wavelength upper-hybrid waves propagating parallel to the
magnetic field, whereas in the strong magnetic field the spectral
energy is transferred to the electrostatic whistlers with oblique
propagation.

 \end{abstract}
 \maketitle
\sloppy
\section{Introduction}
Investigations of various aspects of the beam-plasma instability are
important for our understanding of physical processes occurring in
space and laboratory plasmas. In recent years, regimes of collective
beam-plasma interaction typical to the fast ignition scheme or space
and ionospheric phenomena have attracted particular interest. That
is why most of recent theoretical and numerical studies deals with
strong instabilities excited by counterstreaming electron beams of
comparable densities. Indeed, in the case of isotropic plasma,
results of linear theory relevant for such regimes have been
revisited and dominance of different unstable modes in parameter
space have been determined \cite{bret,bret1}. Nonlinear effects such
as beam trapping and coupling of the Weibel-filamentation
instability with the two-stream instability have been studied in the
framework of the initial-condition problem using both two-
\cite{bret2,fred,kong}  and three-dimensional
\cite{pukh,pukh1,bret3} PIC simulations. As to the case of
magnetized plasma, the particular attention has been focused on
formation and stability of nonlinear phase space structures
(electron holes and tubes) arising during saturation of the
two-stream instability \cite{op,mu,new,sin,umeda,umeda1}.

The goal of this paper is to investigate in details the nonlinear
evolution scenario of a beam-plasma system in the case of low
density beams $n_b\ll n_p$ and to study how such a scenario is
modified by the external magnetic field. Qualitatively, the overall
picture of weak beam-plasma interaction resembles that for a strong
beam, but, in our opinion, more attention should be given to the
study of consecutive energy transfer from one plasma mode to
another. Our interest to this problem is motivated by the fact that
wave-wave interaction due to the beam nonlinearity that is
responsible for the energy exchange between plasma modes in a
uniform beam-plasma system does also play an important role in the
realistic problem of beam injection through a plasma boundary. In
one-dimensional simulations \cite{tim} we have found the regime of
beam relaxation, which is characterized by the regular nonlinear
beam dynamics in electric fields of resonant waves even in a
quasistationary turbulent state. The beam nonlinearity in this
regime results in growth of coherent wave packets instead of waves
with chaotic phases and leads to saturation of the pumping power.
The key argument against feasibility of such a regime is a
one-dimensional character of our simulations. Thus, it is
interesting to study whether nonlinear beam dynamics plays an
essential role in multidimensional problems and whether multimode
character of the beam-plasma instability is able to destroy
correlation effects. For this purpose, we simulate two-dimensional
beam-plasma interaction in the framework of the periodic boundary
problem, in which different resonant modes interact with each other
for a long time due to the beam nonlinearity.

In Sec. \ref{p2} we choose physical parameters of the beam-plasma
system and numerical parameters of our PIC model. In Sec. \ref{p3}
we study the linear stage of the beam-plasma instability and compare
simulation results with theoretical predictions. Sec. \ref{p4}
presents simulation studies for the nonlinear stage of beam-plasma
interaction. Here, we identify the main nonlinear processes
responsible for the energy exchange between different plasma modes
and study the magnetic field effects. In final Sec. \ref{p5} we
discuss our main results.

\section{Simulation parameters}\label{p2}
Let us consider relaxation of an electron beam with the density
$n_b/n_p=0.002$ and the temperature $T_b=10$ eV propagating with the
velocity $v_b/c=0.382$ ($c$ is the speed of light) in a uniform
plasma with the initial electron temperature $T_e=60$ eV.
\begin{figure*}[tbh]
 \bc\includegraphics[width=480bp]{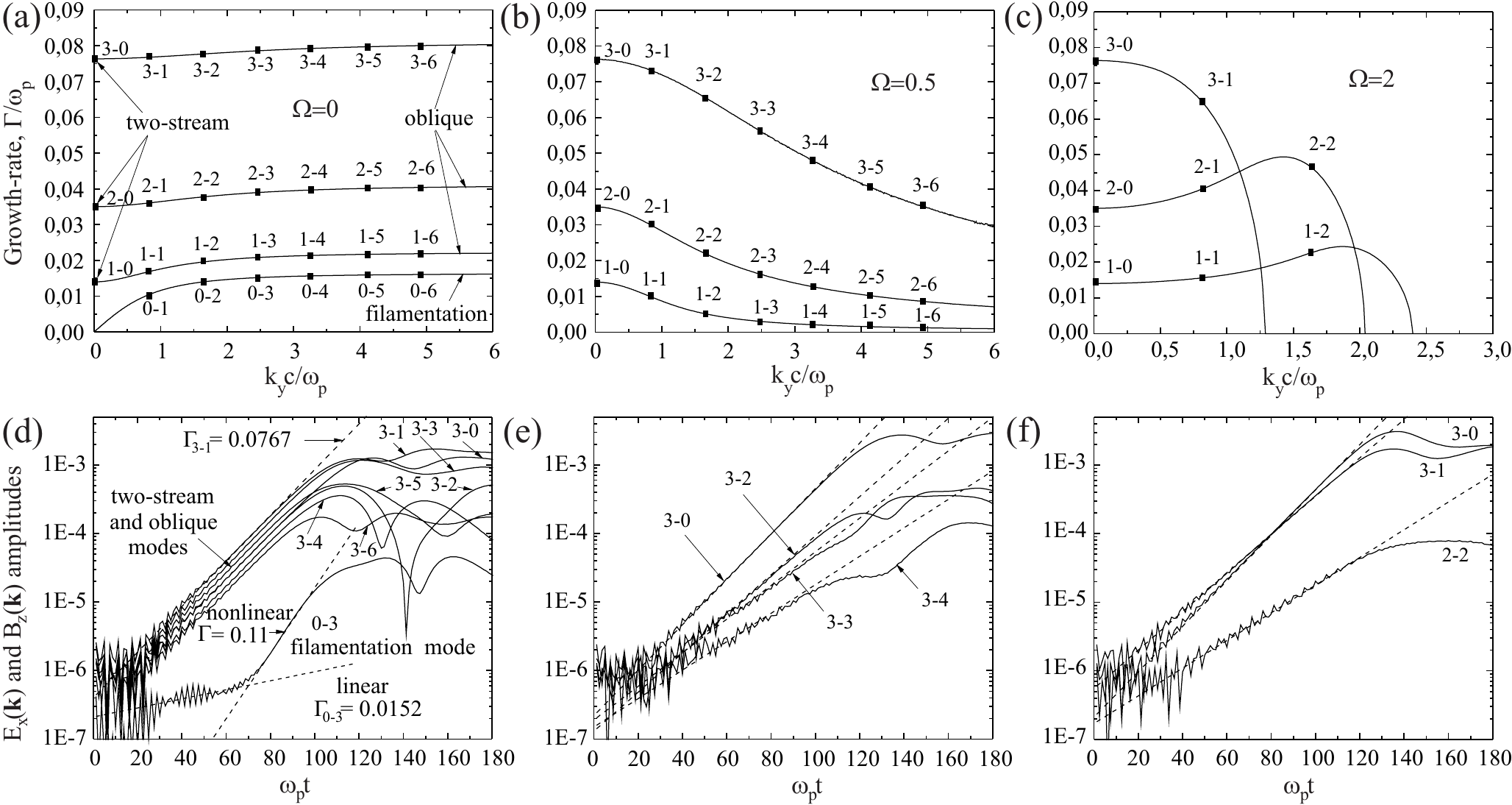} \ec
\caption{(a)--(c) Theoretical growth-rates of unstable modes in the
beam-plasma system for different values of external magnetic field.
(d)--(f) Time-evolution of $E_x$ and $B_z$ Fourier harmonics at the
linear stage of beam-plasma interaction in PIC simulations.
}\label{map}
\end{figure*}
This set of parameters is of a great importance for laboratory
experiments on turbulent plasma heating in mirror traps
\cite{bur,bur1}. Since plasma in such traps is confined by the
external magnetic field $B$ that can be characterized by the ratio
of the cyclotron $\omega_c=eB/(m_e c)$ to the plasma $\omega_p=(4
\pi e^2 n_p/m_e)^{1/2}$ frequency of plasma electrons
$\Omega=\omega_c/\omega_p$, it is interesting to consider different
cases $\Omega=0$, $\Omega=0.5$ and $\Omega=2$ ($e$ and $m_e$ are the
charge and the mass of electrons).

To simulate nonlinear evolution of the beam-plasma system, we use
the standard two-dimensional electromagnetic PIC code with periodic
boundary conditions for particles and fields. Initially, the charge
and current of the electron beam propagating along $x$ are
compensated by charges and currents of plasma electrons. We choose
the simulation box $L_x\times L_y=360 \Delta x \times 384 \Delta y$
and grid sizes $\Delta x=\Delta y=0.02 c/\omega_p$ in such a way as
to fulfill the Cherenkov condition for the modes with wavenumbers
$k_x=6 \pi/L_x=\omega_p/v_b$ and resolve the Debye length. For the
time step we get $\tau=0.01 \omega_p^{-1}$ and for the reasonable
level of noise we use 256 computation particles in each cell. In
addition to beam and plasma electrons, we take into account the
dynamics of ions with the mass $m_i=1836 m_e$.

\section{Linear analysis}\label{p3}
Let us find out what plasma modes should be driven unstable in the
beam-plasma system according to the linear theory and how their
growth-rates depend on the external magnetic field . Due to low beam
and plasma temperatures the linear analysis can be restricted by the
fluid approximation. In this limit, the eigenfrequencies of unstable
plasma oscillations can be found from the equation
\begin{equation}
    \left|k_\alpha k_\beta- k^2
    \delta_{\alpha\beta}+\frac{\omega^2}{c^2}\varepsilon_{\alpha\beta}\right|=0,
\end{equation}
where $\varepsilon_{\alpha\beta}$ is the dielectric tensor with the
components:
\begin{align}
    \varepsilon_{xx} &= 1-\frac{\omega_p^2}{\omega^2}-\frac{n_b}{n_p \gamma^3}\frac{\omega_p^2}{(\omega-k_{x}v_b)^2} \nonumber \\
    &-\frac{n_b}{n_p \gamma} \frac{k_{y}^2 v_b^2}{\omega^2}\frac{\omega_p^2}
    {\left(\omega-k_{x}v_b\right)^2-\omega_c^2/\gamma^2},\nonumber
\\
    \varepsilon_{yy} &= \varepsilon_{zz}=1-\frac{\omega_p^2}{\omega^2-\omega_c^2} \nonumber \\
    &-\frac{n_b}{n_p
    \gamma} \frac{\left(\omega-k_{x}v_b\right)^2}{\omega^2}\frac{\omega_p^2}
    {\left(\omega-k_{x}v_b\right)^2-\omega_c^2/\gamma^2},\nonumber
    \\
    \varepsilon_{xy} &= \varepsilon_{yx}=-\frac{n_b k_{y}v_b \left(\omega-k_{x}v_b\right)\omega_p^2}
    {n_p\gamma \omega^2 \left[ \left(\omega-k_{x}v_b\right)^2-\omega_c^2/\gamma^2\right]},\nonumber
    \\
    \varepsilon_{zx} &= -\varepsilon_{xz}=i\frac{n_b}{n_p
    \gamma^2} \frac{k_{y}v_b}{\omega}\frac{\omega_c}{\omega}\frac{\omega_p^2}
    {\left(\omega-k_{x}v_b\right)^2-\omega_c^2/\gamma^2},\nonumber
    \\
    \varepsilon_{yz} &=
    -\varepsilon_{zy}=-i\frac{\omega_c}{\omega}\frac{\omega_p^2}{\omega^2-\omega_c^2} \nonumber
    \\
    &-i\frac{n_b}{n_p
    \gamma^2} \frac{\left(\omega-k_{x}v_b\right) \omega_c}{\omega^2}\frac{\omega_p^2}
    {\left(\omega-k_{x}v_b\right)^2-\omega_c^2/\gamma^2}.\nonumber
\end{align}
Here, $\gamma$ is the relativistic factor of the beam. In our axis
the beam velocity and the magnetic field are directed along $x$ and
the wavevector has the components ${\bf k}=(k_x,k_y,0)$.

In the spatially periodic system $L_x \times L_y$ possible
wavenumbers are restricted by the following discrete sets:
$$k_x=\frac{2 \pi n}{L_x}=\frac{\omega_p n}{3 v_b},
\quad k_x=\frac{2 \pi m}{L_y}=\frac{\omega_p m}{3 v_b}
\frac{L_x}{L_y},$$ hence, the position of each plasma mode in the
plane $(k_x,k_y)$ can be marked by two integers $n$-$m$. In our
case, the unstable spectrum contains only modes with numbers
$n=0,1,2,3$. Numerical solutions of the dispersion relation for the
growth-rates of these modes as functions of the transverse
wavenumber $k_{y}$ are shown in Fig. \ref{map}(a)-(c). It is seen
that the transition from the case of isotropic plasma with dominant
oblique instabilities to the case of strong magnetic field is
accompanied by the essential reduction in the growth-rates of
oblique modes and complete stabilization of the filamentation
instability.

Let us see how accurately theoretical predictions are reproduced in
PIC simulations. Fig. \ref{map} (d)-(f) show that Fourier harmonics
of $E_x$ and $B_z$ fields does really demonstrate exponential
buildup, but not all of them grow with theoretical growth-rates. In
numerical simulations, dominant modes appear to have an essential
impact on slower instabilities with the same transverse wavenumber.
Indeed, as one can see from Fig. \ref{map} (d), in the case of
isotropic plasma, growth-rates of dominant Langmuir 3-$m$ modes
agree well with theoretical predictions up to large $m$ thus
confirming flow pattern of the observed instability, whereas
instabilities of beam modes 1-$m$ and 2-$m$ are found to be
suppressed. As to the pure transverse filamentary perturbations, it
is shown, using the 0-3 mode as an example, that barely visible
exponential growth of these modes with linear growth-rates switches
rapidly to the nonlinear regime with effective rates exceeding the
maximum growth-rate of the linear theory. Nonlinear nature of this
regime is confirmed by the fact that by the moment $\omega_p t=75$
fluctuations of beam density become comparable in magnitude with the
unperturbed value.

The weak magnetic field $\Omega=0.5$ reduces growth-rates of oblique
upper-hybrid modes 3-$m$ in a good agreement with the linear theory
and keeps beam modes 1-$m$ and 2-$m$ stable. In the strong magnetic
field the spectrum of unstable plasma waves narrows significantly.
It results in the situation when, in addition to the 3-0 and 3-1
modes, we observe the instability of the 2-2 mode. In contrast to
previous cases, this mode now falls in the $k_y$ range, inside which
it becomes dominant.

\section{Nonlinear stage}\label{p4}
\begin{figure}[htb]
 \bc\includegraphics[width=200bp]{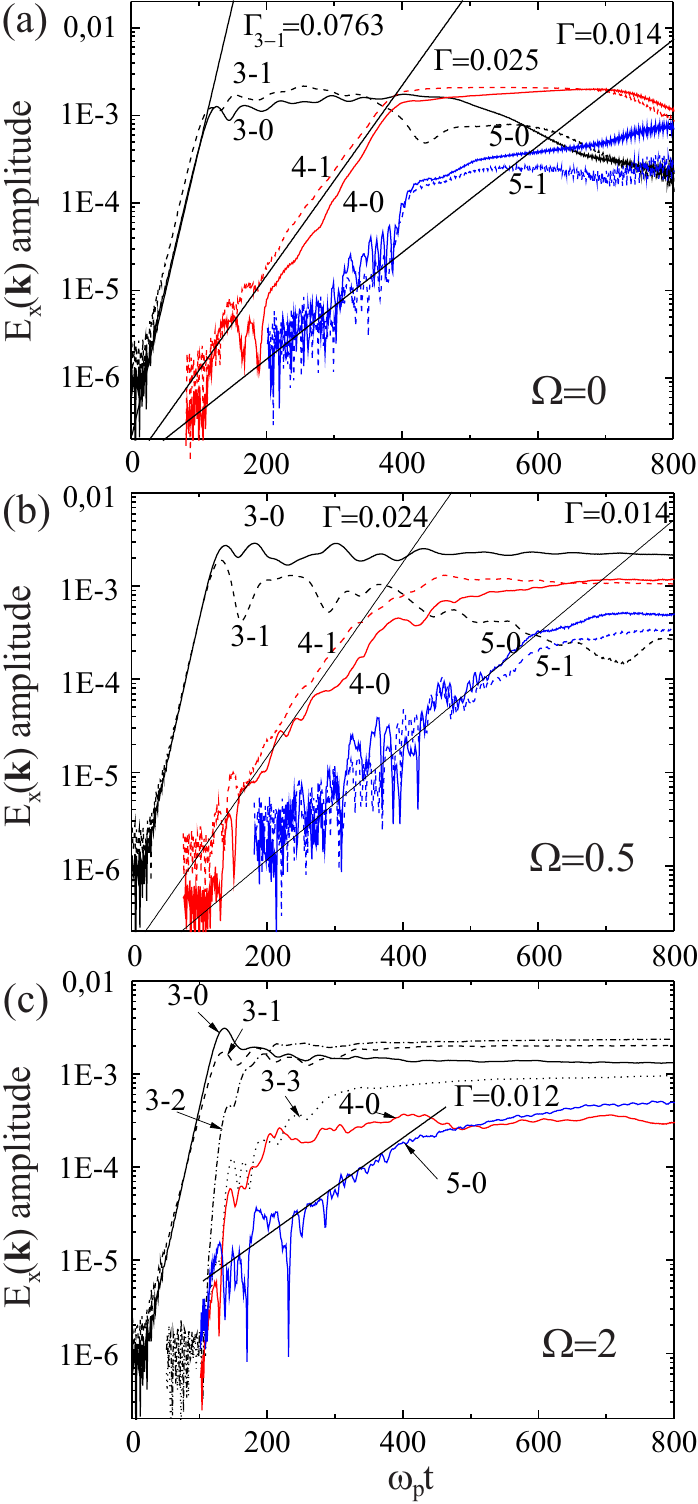} \ec
\caption{(Color online) Time evolution of $E_x$ Fourier harmonics at
the nonlinear stage of beam plasma interaction in different magnetic
fields $\Omega=0$ (a), $\Omega=0.5$ (b) and $\Omega=2$
(c).}\label{waves}
\end{figure}

Before proceeding to the detailed study of nonlinear evolution
scenarios of the beam-plasma system in three different cases
$\Omega=0$, $\Omega=0.5$ and $\Omega=2$, let us briefly describe the
main stages of beam-plasma interaction by considering time evolution
of dominant Fourier harmonics of the electric field $E_x$ presented
in Fig. \ref{waves}.
In the case $\Omega=0$, at the stage of beam trapping, the highest
level of saturation is achieved by modes 3-0 and 3-1. The resulting
nonlinear wave, however, turns out to be unstable against
perturbations with shorter longitudinal wavelengths ($n=4$ and
$n=5$). Among these primarily stable plasma oscillations, the main
role is played by the Langmuir waves with small propagation angles
($m=0,1$).
\begin{figure*}[tbh]
 \bc\includegraphics[width=500bp]{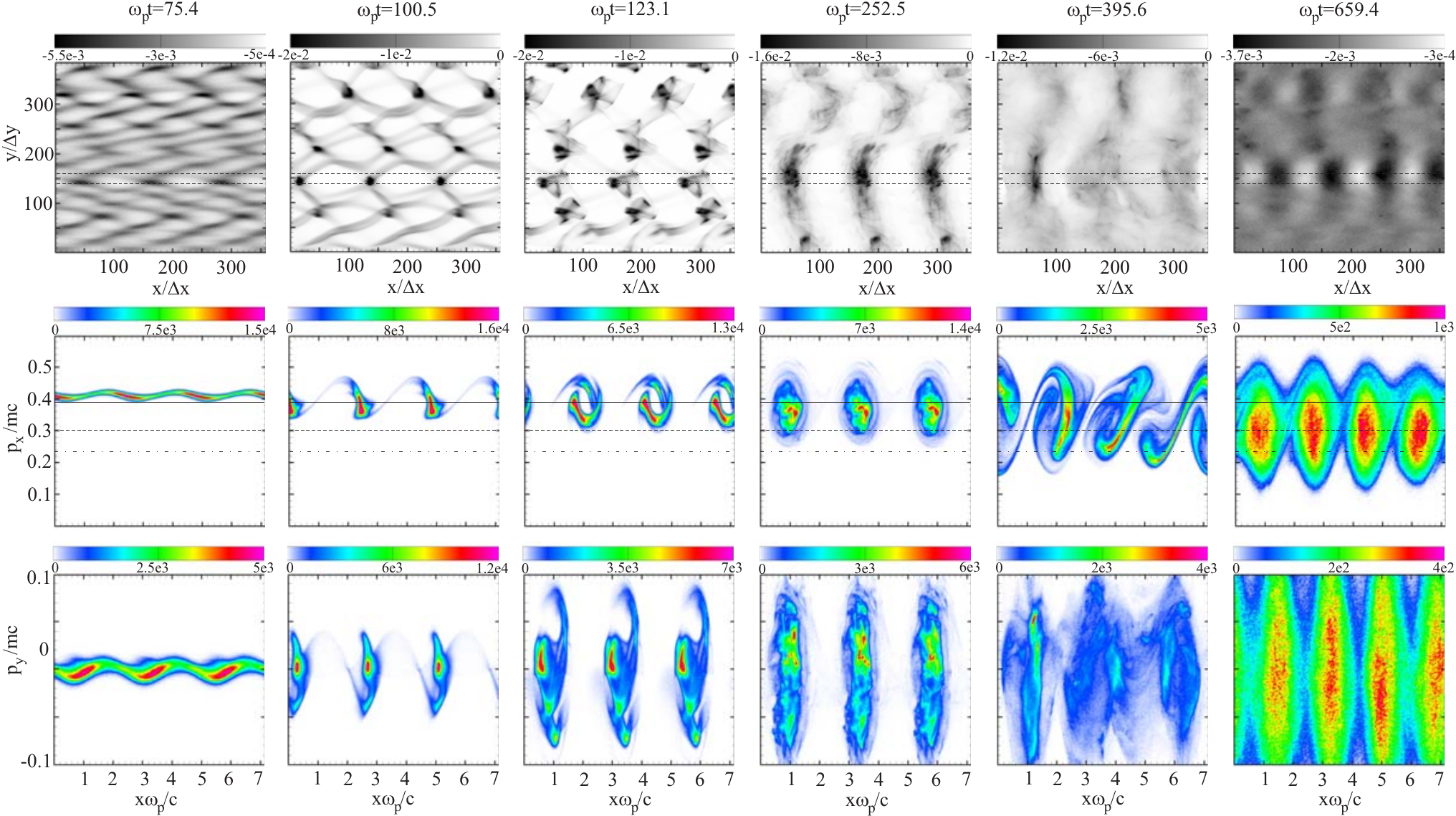} \ec
\caption{(Color online) Evolution of the beam density map (top row)
and beam phase spaces $(x,p_x)$ (middle row) and $(x,p_y)$ (bottom
row) in the case of isotropic plasma. Both phase spaces are obtained
by averaging on $y$ inside the region restricted by horizontal lines
in the beam density map. Lines in the phase space $(x,p_x)$ show
$p_x$ corresponding to phase velocities of different modes: 3-0
(solid), 4-0 (dashed) and 5-0 (dash-dotted). }\label{isotrop}
\end{figure*}
Fig. \ref{waves} (a) shows that the time-averaged growth of these
modes is found to be exponential and saturation levels of secondary
instabilities appear to be so high that from some time the spectrum
of $E_x$ field is dominated by 4-0 and 4-1 modes. Further damping of
these modes is accompanied by the modulation instability resulting
in absorption of their wave energy by plasma electrons.

In the weak magnetic field $\Omega=0.5$, the picture of nonlinear
evolution does not change drastically. Upon saturation of primary
unstable upper-hybrid modes 3-0 and 3-1, we observe the buildup of
secondary instabilities with almost the same averaged growth-rates
as in the previous case (Fig. \ref{waves} (b)). At the saturation
state, however, secondary upper-hybrid modes with $n=4$ do not play
the dominant role. Significant changes in the picture of spectral
energy transfer are found to occur only in the strong magnetic field
$\Omega=2$. In this case, the buildup of initially stable oblique
modes 3-$m$ with $m>1$ becomes more efficient than the growth of
longitudinal oscillations with $n=4$ and $n=5$ (Fig. \ref{waves}
(c)).

\subsection{Isotropic plasma $\bf \Omega=0$}
To study nonlinear beam-plasma interaction in more details, we
present time evolution of the beam density map $n_b (x,y)$ as well
as snapshots of beam phase spaces $(x,p_x)$ and $(x,p_y)$ (Fig.
\ref{isotrop}).

Beam trapping in oblique directions is the first nonlinear
consequence of the prevailing growth of almost electrostatic oblique
modes 3-$m$. Indeed, in the time interval $\omega_p t=75.1\div
100.5$, Fig. \ref{isotrop} demonstrates the process of beam density
modulation under the most unstable oblique modes, which is followed
by the formation of characteristic vortex structures in the phase
space of beam electrons. As we have mentioned, at the beginning of
beam modulation we observe that the filamentation instability
switches to the nonlinear regime. Let us discuss the possible
mechanism of nonlinear excitation of filamentary perturbations.

Modulation of the beam density under unstable modes 3-$m$ and
3-(-$m$) drives density perturbations with the following structure:
$$ \delta n_b \propto e^{\ds i \omega_p (x/v_b - t) +\Gamma_{3-m} t}  \cos (2 \pi y m/L_y).$$
Here, we neglect the difference between the eigenfrequency of the
mode 3-$m$ and the plasma frequency $\omega_p$. Since at the same
time the mode 3-0 perturbs the parallel velocity of the beam
$$ \delta v_b \propto e^{\ds i \omega_p (x/v_b - t) +\Gamma_{3-0} t}, $$
the parallel beam current contains the nonlinear term
$$ \delta j_x \propto \delta n_b \delta v_b \propto e^{\ds (\Gamma_{3-m}+\Gamma_{3-0} )t}  \cos (2 \pi y m/L_y),$$
which generates the magnetic field $B_z$ with the same spatial
structure. Thus, at the early stage, when density and velocity
fluctuations are small enough, filamentary perturbations 0-$m$ due
to the combined effects of two-stream and oblique instabilities
should grow with the approximate rate $ 2\Gamma_{3-0}$. In our
simulations the corresponding growth-rate is slightly lower.
Apparently, it is explained by the fact that growing beam
perturbations at some time cease to be weakly nonlinear.

As it is seen from Fig. \ref{isotrop}, after the trapping stage the
beam breaks up into bunches localized in space in both parallel and
perpendicular directions. One can also see that these bunches are
arranged one behind the other thus forming current layers along the
magnetic field. Since the beam acquires the large spread in
transverse momentum under oblique modes, different current layers
exchange beam particles and merge into one by the moment $\omega_p
t=252.5$. At the same time, in phase spaces $(x,p_x)$ and $(x,p_y)$
we observe the formation of nonlinear BGK-wave, the amplitude of
which, in contrast to the one-dimensional problem, is localized in
the transverse direction.

Let us investigate stability of the resulting nonlinear wave. As we
have mentioned, at this stage, the beam-plasma system is unstable
against short-wavelength almost longitudinal Langmuir oscillations
with $n=4$ and $n=5$. The exponential growth of secondary
instabilities indicates that the initial stage of these
instabilities can be described by the theory linearized in the
amplitudes of unstable perturbations. In this case, however, we
cannot consider a uniform beam-plasma system even with the increased
momentum spread of the beam as an unperturbed state. Excitation of
secondary plasma modes differs markedly from Cherenkov mechanism.
Fig. \ref{waves} and \ref{isotrop} show that the instability of the
mode 5-0, for example, starts to grow when there are no any
particles in resonance with this mode (the momentum corresponding to
the phase velocity of this mode is shown in the phase space
$(x,p_x)$ as the dash-dotted line). The stability analysis of a
one-dimensional (uniform along $y$) nonlinear BGK-wave as an exact
stationary solution of the Vlasov-Poisson system seems to be a more
adequate theoretical approach in our case. According to the theory
proposed in Ref. \cite{kruer,gold}, efficient energy exchange
between a plasma mode with the frequency $\omega$ and the wavenumber
$k$ and a nonlinear BGK-wave with the phase velocity $v_0$ and the
bounce-frequency $\omega_b$ occurs if the following resonance
condition is fulfilled: $\omega=k v_0+N\omega_b$. Let us compute
growth-rates of sideband modes 4-0 and 5-0 driven unstable due to
resonances $N=-1$ and $N=-3$, respectively.

\begin{figure}[htb]
 \bc\includegraphics[width=210bp]{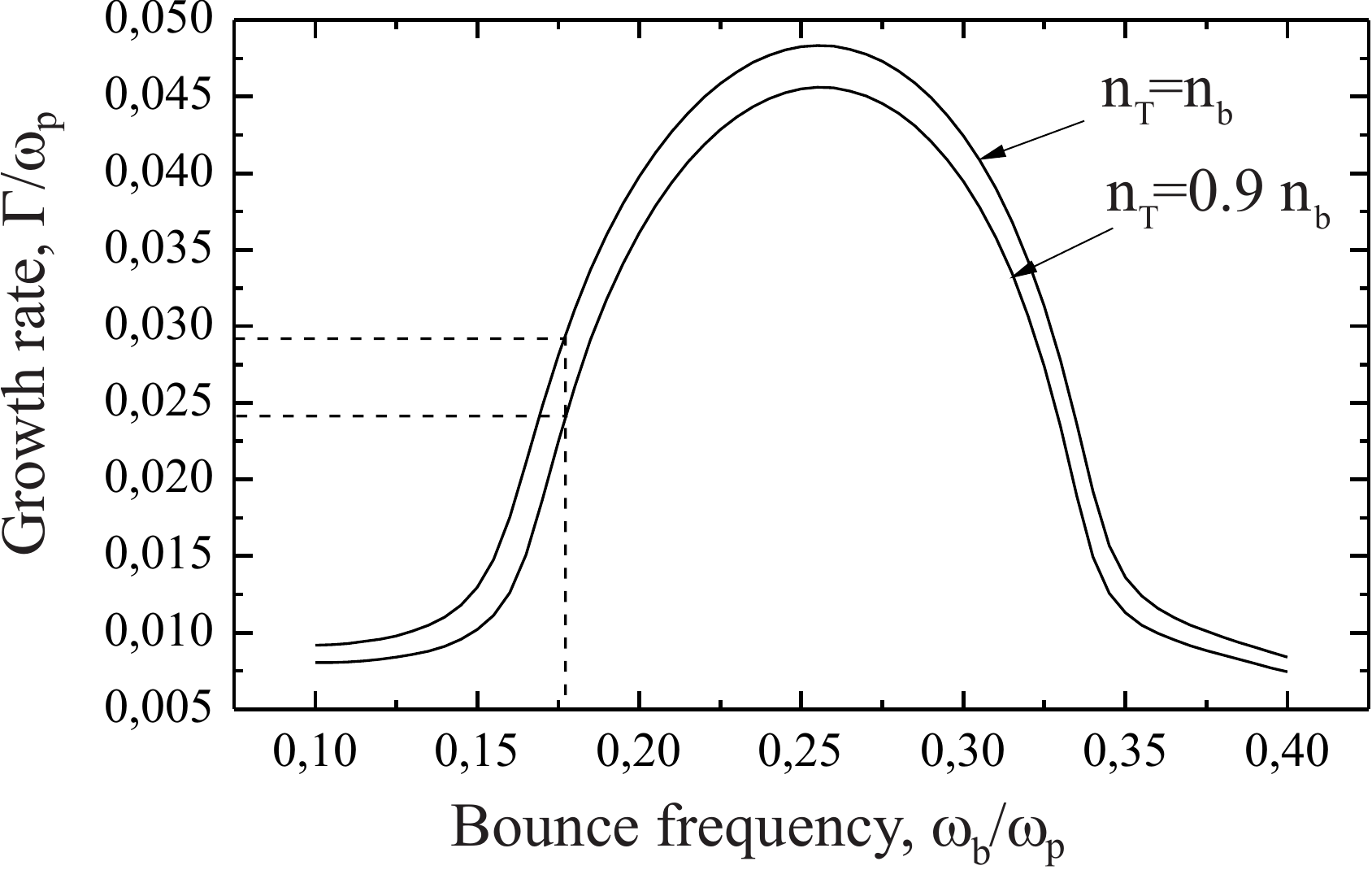} \ec
\caption{Dependence of the growth-rate of sideband instability for
the mode 4-0 on the bounce frequency of trapped electrons in a
one-dimensional BGK-wave. }\label{side}
\end{figure}

For this purpose, we simplify the general theory of Ref. \cite{gold}
using the following assumptions. Most of beam particles is
concentrated near the bottom of the potential well with the
parabolic profile $V(x)=-e \varphi_0 (1-k_0^2 x^2/2)$ and their
distribution function $f(W)$ is constant inside the energy range $-e
\varphi_0<W<-0.9 e \varphi_0=W_0$ ($\varphi_0$ denotes the amplitude
of electrostatic potential in the BGK-wave). In this case, all beam
particles participate in bounce oscillations with the same frequency
$\omega_b=k_0 (e \varphi_0/m)^{1/2}$ and their maximum deviation
length  from the center of the potential well is determined by
$$ a=\frac{1}{k_0} \sqrt{2 \left(\frac{W_0}{e \varphi_0}+1\right)}.$$
Goldman theory \cite{gold} results in the following dispersion
relation
\begin{equation}\label{e2}
    \left(\varepsilon(k,\omega)+\chi_{0,0}\right) \left(\varepsilon(k-2k_0,\omega-2\omega_0)
    +\chi_{-2,-2}\right) =\chi_{0,-2}^2,
\end{equation}
where
$$ \chi_{l,s}=-\omega_p^2 \frac{n_T}{n_p} \sum\limits_{N=1}^{\infty} \frac{4 N^2 J_N(z_l) J_N(z_s)}{z_l z_s \left((\omega-k v_0)^2-N^2 \omega_b^2\right)},$$
$\varepsilon (k,\omega)$ is the linear plasma permittivity, $J_N$
are the Bessel functions of arguments $z_j=(k+j k_0)a$ and $n_T$ is
the space-averaged density of trapped particles. In order to
calculate the growth-rate of the mode 4-0 as the function of the
bounce frequency of BGK-wave, we should take into account the
contribution of only one resonance $N=1$. Numerical solution of such
a dispersion relation is presented in Fig. \ref{side}.
\begin{figure*}[tbh]
 \bc\includegraphics[width=450bp]{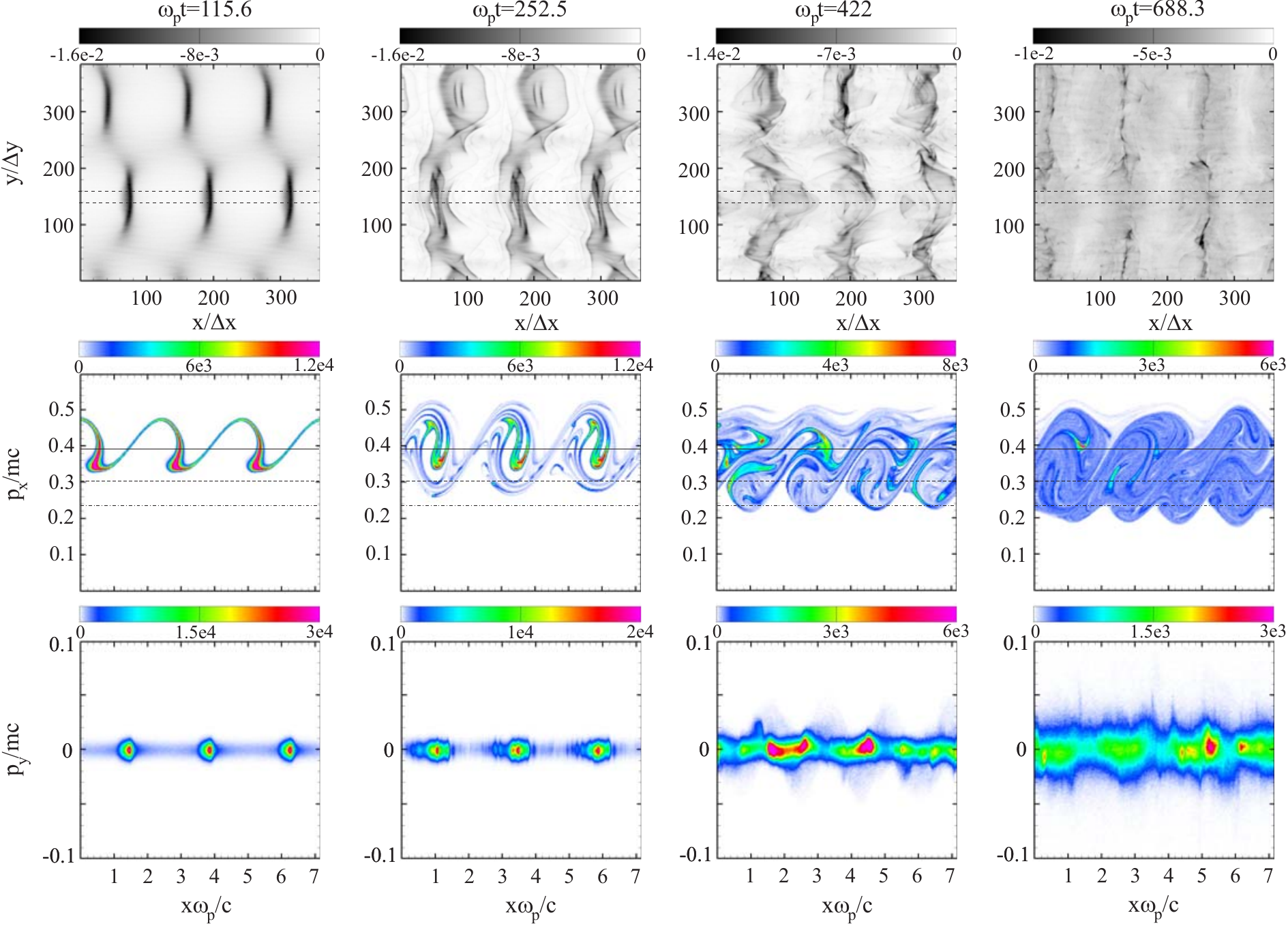} \ec
\caption{(Color online) Evolution of the beam density map (top row)
and beam phase spaces $(x,p_x)$ (middle row) and $(x,p_y)$ (bottom
row) in the case $\Omega=0.5$. Both phase spaces are obtained by
averaging on $y$ inside the region restricted by horizontal lines in
the beam density map. Lines in the phase space $(x,p_x)$ show $p_x$
corresponding to phase velocities of different modes: 3-0 (solid),
4-0 (dashed) and 5-0 (dash-dotted).}\label{m05}
\end{figure*}
From this figure we notice that for the observed amplitude of the
BGK-wave the best agreement between theoretical and simulation
results is achieved, if the effective density of trapped particles
is slightly lower than the beam density: $n_T=0.9 n_b$. If for the
same amplitude of nonlinear wave we take into account the
contribution of the resonance $N=3$ in the dispersion equation, we
are also able to predict the growth-rate of the mode 5-0, which is
found to reach the value $\Gamma/\omega_p \simeq 0.015$. Besides
longitudinal modes 4-0 and 5-0, which can be predicted
theoretically, we also observe instabilities of oblique modes 4-$m$
and 5-$m$ with small propagation angles.

As to the nonlinear saturation of sideband instabilities, in the
time interval $\omega_p t=395.6\div 659.4$ (Fig. \ref{isotrop}) this
process is accompanied by transformation of the primary BGK-wave
with $n=3$ to the other nonlinear wave with $n=4$. From the motion
of beam particles in the phase space $(x,p_x)$ we conclude that the
wave energy comes to the secondary mode 4-0 not only from the
kinetic energy of the beam that is decelerated by this slower wave,
but also from the primary unstable mode 3-0 that loses most of its
energy during nonlinear wave-wave interactions. In the later stage,
plasma nonlinearities come into force and the BGK-wave with $n=4$ is
dissipated due to the local (in $y$) buildup of the modulation
instability.

\subsection{Weak magnetic field $\bf \Omega=0.5$}
In the weak magnetic field, due to decreasing in growth-rates of
oblique instabilities, the main role at the nonlinear stage of
beam-plasma interaction is played by modes 3-0 and 3-1. The
exponential growth of these modes is saturated by beam trapping. In
the chosen local region, this nonlinear process can be considered as
one-dimensional. Indeed, Fig. \ref{m05} shows that at the moments
$\omega_p t=115.6$ and $\omega_p t=252.5$ beam particles mix inside
the trapping phase-space region of the 3-0 mode without a visible
increase in their transverse momentum.

Due to further damping of the primary mode 3-1, we observe the
formation of almost one-dimensional BGK-wave. Since the amplitude of
this nonlinear wave is comparable with that in the previous case,
secondary instabilities of almost longitudinal sideband modes with
$n=4$ and $n=5$ grow with the same growth-rates as in the isotropic
plasma [Fig. \ref{waves}(b)]. Nonlinear saturation of these modes,
however, is not accompanied by the complete wave energy transfer
from primary unstable modes to sideband modes. As one can see from
Fig. \ref{m05} demonstrating snapshots of the phase space $(x,p_x)$
at the moments $\omega_p t=422$ and $\omega_p t=688.3$, beam
particles move simultaneously in fields of both primary ($n=3$) and
secondary modes ($n=4$).
\begin{figure*}[tbh]
 \bc\includegraphics[width=460bp]{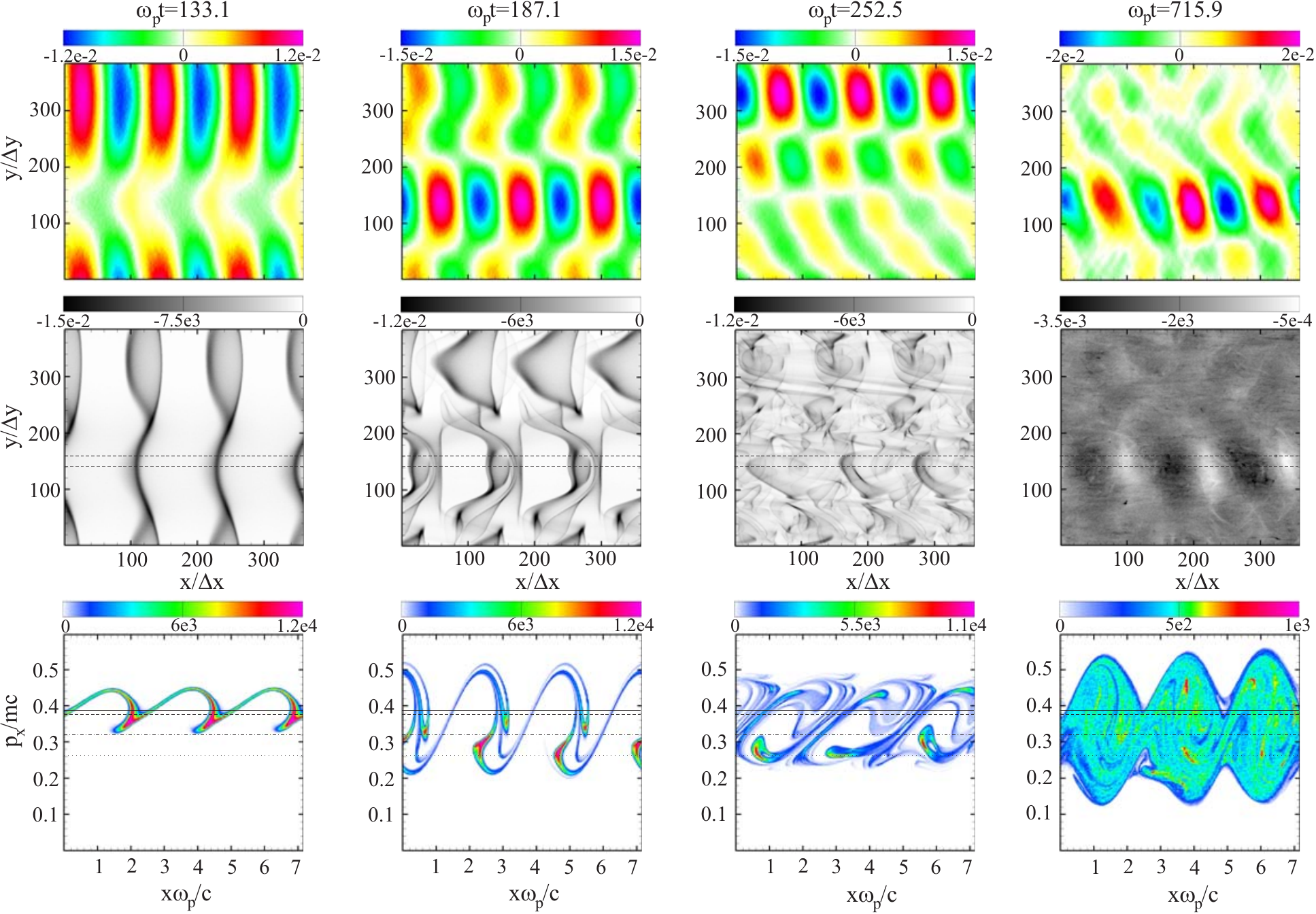} \ec
\caption{(Color online) Evolution of the $E_x$ map (top row), the
beam density map (middle row) and the phase space of the beam
$(x,p_x)$ (bottom row) in the case $\Omega=2$. The phase space is
obtained by averaging on $y$ inside the region restricted by
horizontal lines in the beam density map. Lines in the phase space
$(x,p_x)$ show $p_x$ corresponding to phase velocities of different
modes: 3-0 (solid), 3-1 (dashed), 3-2 (dash-dotted) and 3-3
(dotted).}\label{m2}
\end{figure*}
The beam spread in transverse momentum, increasing due to
excitation of oblique sideband modes 4-1 and 4-2, is found to be
much smaller than in the case of isotropic plasma.

Thus, in the weakly magnetized plasma the nonlinear stage of the
beam-plasma instability starts with almost one-dimensional beam
trapping. The resulting BGK-wave propagating parallel to the
magnetic field drives sideband instabilities of various upper-hybrid
modes. In contrast to the previous case, nonlinear interactions
between primary and secondary modes do not result in the substantial
energy transfer and dominant modes 3-0 and 4-0 have comparable
amplitudes in the quasi-stationary state.

\subsection{Strong magnetic field $\bf \Omega=2$}
In the strong magnetic field the unstable spectrum narrows
significantly. Beam trapping continues to be the main nonlinear
process resulting in saturation of dominant unstable modes 3-0 and
3-1. The contribution of oblique mode  becomes herewith more
pronounced. In Fig. \ref{m2} demonstrating snapshots of the electric
field map $E_x(x,y)$, the beam density map $n_b(x,y)$ and the beam
phase space $(x,p_x)$, one can see that trapping of beam particles
in different regions along $y$ starts at different moments of time.
By the moment $\omega_p t=187.1$, the nonlinear BGK-wave has not yet
been formed, but primarily stable modes 3-2, 3-3 and 4-0 have
already grown up to large amplitudes. All of these modes get into
the Cherenkov resonance with beam particles, but their rapid
intermittent growth does not allow to clarify unambiguously what
excitation mechanism (Cherenkov or nonlinear) plays the main role.

Fig. \ref{waves}(c) shows that, upon saturation of secondary
unstable modes, the wave spectrum is dominated by 3-$m$ modes with
the prevailing role played by the oblique mode 3-2. As to determine
the branch of plasma oscillations, to which secondary modes with
$m\geq 2$ belong, we should find real parts of their frequencies.
This can be done by using the relationship between different field
components $E_x$, $E_y$ and $B_z$:
$$ \omega({\bf k})= \frac{c}{B_z({\bf k})}\left(k_x E_y({\bf k}) - k_y E_x({\bf k})\right). $$
Substituting simulation results into this formula, we identify
secondary modes as electrostatic whistler waves.

Dominance of plasma modes with $n=3$ is also visible in the phase
space of the beam. In Fig. \ref{m2} at the moment $\omega_p
t=252.5$, typical vortex structures arising from trapping of beam
particles by different 3-$m$ modes are observed clearly. Due to the
difference in parallel phase velocity, trapping regions of these
modes move relative to each other, but when bottoms of potential
wells have the same $x$-coordinates, electric fields of excited
modes are summed coherently and form the large potential well
resulting in phase space mixing of beam particles from different
trapping regions. From spatial structures of the electric field and
the beam density at the moments $\omega_p t=252.5$ and $\omega_p
t=715.9$ we see that the spatial region of coherent wave-wave
interaction is localized in the transverse direction and moves along
$y$. Intense phase-space mixing of beam particles inside this region
makes the nonlinear interaction between plasma modes with the same
parallel wavenumbers $k_{\parallel}$ to be more efficient than
interaction between modes with different $n$.

Thus, in strongly magnetized plasmas, nonlinear wave-wave
interaction via shared trapped particles is marked by the prevailing
energy transfer from linearly unstable modes to initially stable
oblique whistlers with the same parallel wavenumber. In spite of
excitation of oblique plasma waves, the beam spread in transverse
momentum remains unchanged for the whole period of beam-plasma
interaction.

\section{Discussion of simulation results}\label{p5}
Detailed two-dimensional PIC simulations show that collective
interaction of a cold low-density electron beam with an isotropic
plasma evolves according to the following scenario. Prevailing
growth of oblique linear instabilities is followed by trapping of
beam particles in oblique directions and substantial spreading of
the beam in transverse momentum. The linear growth-rate of the
filamentation instability in our case is much smaller than
growth-rates of oblique and two-stream instabilities, but from the
certain time filamentary perturbations demonstrate rapid growth due
to nonlinear interaction of primary unstable modes. Simultaneous
buildup of two-stream, oblique and filamentation instabilities
results in the situation when the beam appears to be divided into
separate bunches localized spatially in both longitudinal and
transverse directions. Current layers constructed from these bunches
merge into one thus forming the nonlinear BGK-wave localized in $y$.
This nonlinear equilibrium is found to be unstable against
oscillations with frequencies, which in the reference frame of
BGK-wave equal to the harmonics of the bounce frequency of trapped
electrons. Growth-rates of secondary sideband modes propagating
parallel to the beam direction are well predicted by the
one-dimensional theory. Saturation of sideband instabilities is
characterized by almost complete energy transfer from primary
unstable modes to almost longitudinal Langmuir modes with shorter
wavelengths. It results in the formation of the BGK-wave with
$k_{\parallel}>\omega_p/v_b$, which is followed by dissipation of
its energy due to the modulation instability.

In the weak magnetic field, oblique instabilities cease to dominate
in the linearly unstable spectrum and the nonlinear stage of
beam-plasma interaction starts with almost one-dimensional beam
trapping followed by the formation of one-dimensional BGK-wave. In
contrast to the case of isotropic plasma, the beam does not acquire
significant spread in transverse momentum at this stage. Although
sideband instabilities of short-wavelength modes 4-$m$ and 5-$m$
grow with the same rates as in the isotropic plasma, saturation
levels of these secondary upper-hybrid modes appear to be lower. In
this case, the prevailing energy transfer at the saturation stage
remains to be directed along $k_{\parallel}$, but nonlinear
interaction between primary and secondary modes becomes less
efficient.

The strong magnetic field does not affect essentially the initial
stage of beam trapping by linearly unstable modes with small
propagation angles, but modifies substantially the spectrum of
secondary modes arising due to nonlinear wave-wave interaction via
the beam nonlinearity. In this case, the wave energy of primary
unstable modes is transferred to oblique modes with the same
parallel wavenumber. It means that the energy flow is directed
predominantly along $k_{\perp}$.
\begin{figure}[htb]
 \bc\includegraphics[width=190bp]{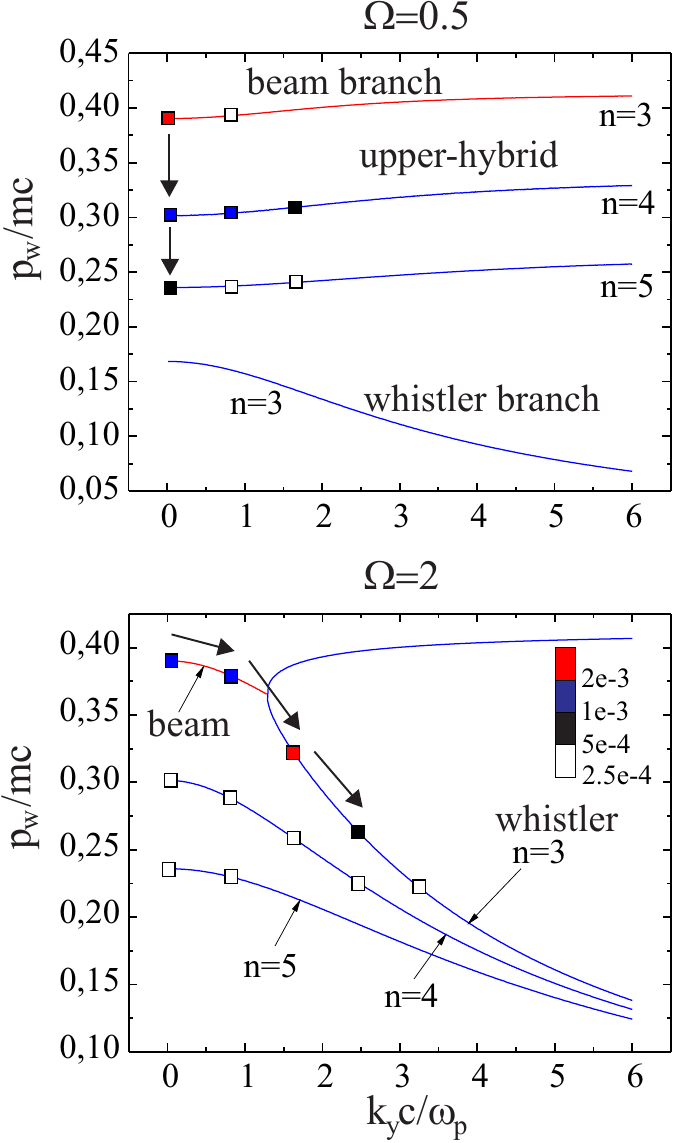} \ec
\caption{(Color online) Phase velocities of linearly stable (blue)
and linearly unstable (red) modes in the beam-plasma system in terms
of momentum of resonant electrons. Bold arrows indicate the
prevailing direction of energy flow.}\label{disp}
\end{figure}

In order to understand why the direction of spectral energy transfer
at the nonlinear stage of beam-plasma interaction changes
drastically with the increase of external magnetic field, let us
consider some dispersion characteristics of plasma modes that get
into the Cherenkov resonance with the trapped electron beam. Let us
express phase velocities of these modes in terms of momentum of
resonant electrons
$$\frac{p_{\mbox{w}}}{m_e c}=\left(\left(\frac{k_{\parallel}c}{\omega_{{\bf k}}}\right)^2-1\right)^{-1/2}.$$
Fig. \ref{disp} presents the dependence of this momentum on the
transverse wavenumber for different branches of plasma waves and
different magnetic fields. Squares show positions of modes that can
be resolved in our periodic system and colors indicate amplitudes of
these modes at the end of simulation run. The main difference
between cases $\Omega=0.5$ and $\Omega=2$ is that in the weak
magnetic field the unstable beam branch couples to the upper-hybrid
branch whereas in the strong magnetic field it couples to the
whistler branch. In the former case the phase velocity of unstable
plasma modes with the same parallel wavenumber increases with the
increase of $k_{\perp}$, whereas in the latter case it decreases
even after the transition from unstable beam modes to stable
whistlers. Thus, the observed distribution of wave energy between
different plasma modes can be explained, if nonlinear wave-wave
interactions due to the beam nonlinearity are governed by the
following rules: (i) interaction between modes is more efficient, if
these modes have the same $k_{\parallel}$; and (ii) wave energy is
transferred to modes with lower phase velocities. It explains why,
in the weak magnetic field, oblique unstable modes 3-$m$, despite of
their large growth-rates, are saturated at rather low levels  and
why the mode 3-1 loses its energy at the later nonlinear stage. In
the absence of slower modes with $k_{\parallel}=\omega_p/v_b$, wave
energy here is transferred to the modes with
$k_{\parallel}>\omega_p/v_b$ arising due to sideband instabilities.
In the strong magnetic field, the efficient energy transfer from
primary unstable modes 3-0 and 3-1 to other modes with the same
$k_{\parallel}$ and lower phase velocities becomes possible. That is
why the spectrum of plasma oscillations in the relaxed beam-plasma
system in this case is dominated by oblique whistler modes.

This work is supported by President grant NSh-5118.2012.2,  grant
11.G34.31.0033 of the Russian Federation Government and RFBR grants
11-02-00563-a, 11-01-00249-a.

\end{document}